\documentclass[12pt,a4paper]{article}
\usepackage{graphicx}
\usepackage{times}
\title{\bf Photometric monitoring of Luminous Blue Variables}
\author{Carla Buemi$^1$, Elisa Distefano$^{1,2}$, Paolo Leto$^1$, Francesco Schillir{\`o}$^3$, Corrado Trigilio$^1$, \\Grazia Umana$^1$, Stefano Bernabei$^4$, Giuseppe Cutispoto$^1$, and Sergio Messina$^1$\\
\vspace{1cm}\\
\normalsize $^1$ INAF-Osservatorio Astrofisico di Catania, Catania Italy\\ 
\normalsize $^2$Dipartimento di Fisica e Astronomia, Universit\'a di Catania, Catania Italy\\
\normalsize $^3$ INAF- Istituto di Radioastronomia, Noto, Italy\\
\normalsize $^4$ Osservatorio astronomico, Bologna,Italy}
\date{\mbox{}}
\begin{document}
\maketitle
\pagestyle{empty}
%
% WE REDEFINE THE plain LaTeX PAGESTYLE !!! 
% THIS PAGESTYLE WILL BE USED FOR THE FIRST PAGE ONLY !
%
\def\bull{\vrule height .9ex width .8ex depth -.1ex}
\makeatletter
\def\ps@plain{\let\@mkboth\gobbletwo
\def\@oddhead{}\def\@oddfoot{\hfil\tiny\bull\quad
``The multi-wavelength view of hot, massive stars''; 39$^{\rm th}$ Li\`ege Int.\ Astroph.\ Coll., 12-16 July 2010 \quad\bull}%
\def\@evenhead{}\let\@evenfoot\@oddfoot}
\makeatother
%
% AND DEFINE OUR MACROS FOR THE REFERENCE LIST
% I.E \beginrefer \refer and \endrefer
%
\def\beginrefer{\section*{References}%
\begin{quotation}\mbox{}\par}
\def\refer#1\par{{\setlength{\parindent}{-\leftmargin}\indent#1\par}}
\def\endrefer{\end{quotation}}
%
% BEGIN THE ABSTRACT CHAPTER WITH \noindent\small, ENCLOSE IT IN A GROUP
% AND BOLDFACE THE TITLE.
%
{\noindent\small{\bf Abstract:} 
%We present the first results of our observational program for the study of 
We present some preliminary results from our program of intensive near-infrared photometric monitoring of 
 a sample of confirmed and candidate Luminous Blue Variables
(LBVs) conducted from 2008 to 2010.  Clear long-term variability has been observed for Wray~17-96 
and V481~Sct, with overall brightness variation greater than 1 mag in the J band. Other sources, such as
LBV~1806-20 showed detectable variability
with amplitudes of few tenths of a magnitude with time-scale of about 60 days. 

}
%
% NOW COMES THE MAIN BODY OF THE ARTICLE
%
\section{Introduction}
The class of Luminous Blue Variables (LBVs) consists of luminous and massive
stars, that are believed to go through a short 
but violent transition phase of evolution from the main sequence towards the Wolf-Rayet stage
(Humphreys \& Davidson, 1994; Langer et al., 1994).
LBVs are well know to show a combination of spectral and photometric 
variability (S-Doradus type variability and/or $\eta$-Car type eruptions), 
whose origin is not yet well understood, 
despite the crucial role that such objects play in the stellar evolution
of massive stars (van~ Genderen, 2001; Kotak \& Vink, 2006). 
A detailed analysis of the time scale variability of LBV stars can 
provide useful insights in the understanding the evolution of such
objects and in the knowledge of the physical mechanisms that trigger the great 
giant eruption. Some objects, like $\eta$~Car, P~Cyg and AG~Car,
has been extensively monitored, but often fragmentary observations exists for others.
We are thus conducting a long term multiwavelength photometric monitoring
of a sample of confirmed and candidate LBVs. 
We present the most interesting results of the first two years of observations,
performed by using the REMIR
infrared imaging camera available at the REM (Rapid Eye Mount) telescope.

\section{Observations and results}
Here, we present observations taken with the 60 cm robotic Rapid Eye Mount (REM) telescope, located at the European Southern Observatory (ESO) (La Silla, Chile). Data were acquired in the V, R, I, J, H and K bands since 2008 April up to now with a time sampling of one measure per week, as a part of a multiwavelength (optical and infrared) monitoring campaign of 25 LBVs and cLBVs, listed in Table 1,dedicated to investigate their photometric variability. 
\begin{table}
\caption{List of targets. The J magnitude are from the 2MASS catalog.}
\label{list}
\small
\begin{center}
\begin{tabular}{ |l| l l r|l| l l r|}
\hline
Name & $\alpha$(2000) & $\delta$(2000)  & J mag & Name & $\alpha$(2000) & $\delta$(2000)  & J mag  \\
\hline
PV Vel       & 09:15:54.8 &$-$49:58:24.6 &{\it 4.44}& V4375 Sgr     & 17:48:14.0 &$-$28:00:53.1 &{\it 4.82}\\	
HR Car       & 10:22:53.8 &$-$59:37:28.4 & 4.56     & LBV 1806$-$20 & 18:08:40.3 &$-$20:24:41.1 & 13.66\\ 	 
GSC08958     & 10:53:59.6 &$-$60:26:44.3 & 7.32     & V4029 Sgr     & 18:21:14.9 &$-$16:22:31.8 & 4.60\\
AG Car       & 10:56:11.6 &$-$60:27:12.8 & 5.42     & V4030 Sgr     & 18:21:19.5 &$-$16:22:26.1 &{\it 5.14}\\
V432 Car     & 11:08:40.1 &$-$60:42:51.7 & 7.96     & V481 Sct      & 18:33:55.3 &$-$06:58:38.7 & 8.36\\
Sher 25      & 11:15:07.8 &$-$61:15:17.6 & 8.60     & GAL025        & 18:37:05.2 &$-$06:29:38.0 &{\it 15.80}\\
W243         & 16:47:07.5 &$-$45:52:29.2 & 6.41     & V452 Sct      & 18:39:26.1 &$-$13:50:47.1 & 7.86\\
$\zeta$ Sco  & 16:53:59.7 &$-$42:21:43.3 & 3.59     & GAL026        & 18:39:32.2 &$-$05:44:20.5 & 8.00\\
V1104 Sco    & 17:06:53.9 &$-$42:36:39.7 & 6.71     & V1672 Aql     & 19:00:10.9 &$+$03:45:47.1 & 12.16\\
Wray17$-$96  & 17:41:35.4 &$-$30:06:38.8 & 6.71     & V1429 Aql     & 19:21:34.0 &$+$14:52:56.9 & 6.09\\  	
V905 Sco     & 17:41:59.0 &$-$33:30:13.7 & 3.55     & W 51          & 19:23:42.3 &$+$14:30:33.0 &{\it 14.32}\\	
V4650 Sgr    & 17:46:18.0 &$-$28:49:03.5 & 12.31  	& V1302 Aql     & 19 26 48.3 &$+$11 21 16.7 & 5.47\\
WR102ka      & 17:46:18.1 &$-$29:01:36.6 & 12.98    & & & & \\ 
\hline   
\end{tabular}
\end{center}
\end{table}
 As the monitoring program is still ongoing, we present just the most interesting, although partial, results from the infrared light curves. Since our goal was to investigate variability of these objects, we carried out differential photometry of the candidates with respect to non-variable comparison stars in each target field, using the ensemble photometry technique (Gilliland \& Brown 1988; Everett \& Howell 2001) implemented in the 
software  {\it ARCO} (Automatic Reduction of CCD Observation, developed by Distefano et al. (2007). This software allows to select the  stars for the ensemble among those in common to all frames, iteratively rejecting stars that were found to have either a systematic variation in the instrumental magnitude or large errors. The NIR magnitudes were calibrated against  magnitude in the 2MASS catalog. 

The accuracy and time resolution of the monitoring allow us to detect even small amplitude variability. Photometric errors, as derived following the standard method described by Everett \& Howell (2001),
 are about 0.01 mag on most of the individual measurements; thus the photometric data quality is good enough to distinguish photometric variation at close to 0.05 mag level.

Flux curves and observed color variations are presented for the clearest cases of both long and/or short term detected variability. 

\subsection{V481~Sct}
Strong variation in amplitude has been observed in the light curves of this source (Fig. 1).  The IR variability is large ($\Delta$J$\approx$1.0) and the rise seems to be faster between the first two runs, although the gap in the coverage does not allow us to follow the entire trend. The star further brightens  up to a total magnitude variation of $\Delta$J$\approx$1.2 on a time scale of about two years. V481~Sct experienced a similar photometric variation between 2003-2004, as reported by Clark et al. (2005), suggesting a sort of cyclic photometric behavior.  Smaller ($\Delta$J$\approx$0.3) and shorter (time scale of about 60 days) flux variations seems to be superimposed to the long term brightening. 
The infrared colors do not show a systematic trend but exhibit a quite irregular change, making it hard to derive the mechanism of the variability on the basis of the near-IR photometry alone. We note that the mean J-K and J-H seem to decrease during the first stellar brightening, suggesting a stellar blueing. However, the trend changes during the third run, suggesting that more than one process is acting. 

\begin{figure}[h]
\label{v481}
\centering
\includegraphics[width=15cm]{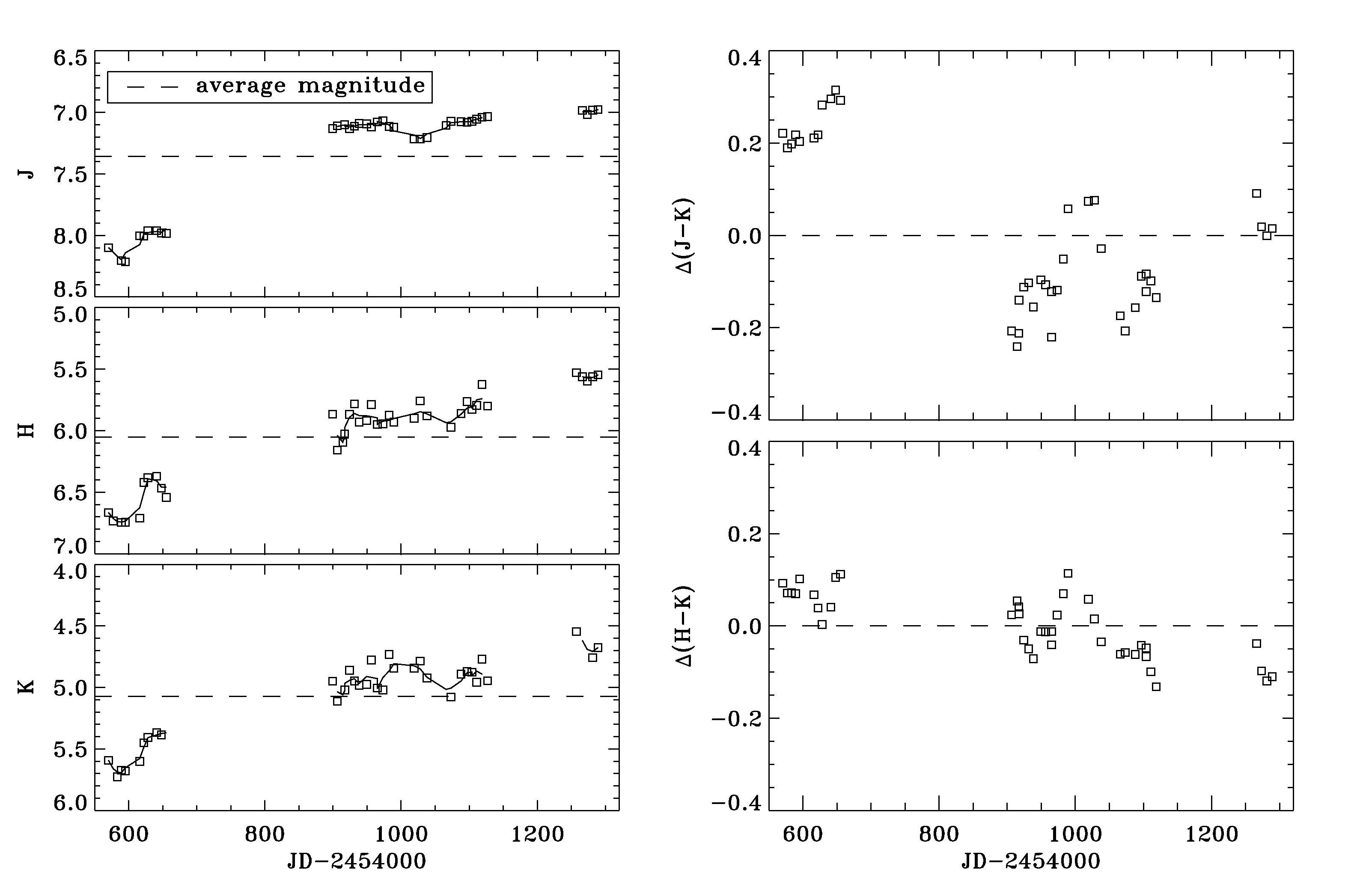}
\caption{Light and color curves for V481~Sct. The solid lines represent the smoothed brightness.}
\end{figure}
\subsection{Wray~17-96}
As for V481~Sct, the brightness of Wray~17-96 varies significantly over the entire interval of our observations (Fig. 2). The total variation  amplitudes tend to decrease with wavelength, spanning from 1~mag in the K band, to 1.45~mag in the J band, but clear small amplitude ($\Delta$K$\approx$0.5) variability can be observed  also on shorter time scales of about 60 days.  
As shown in the Figure 3, the infrared color became bluer as the flux rises, from the beginning to the end of our observations, indicating an increasing of stellar temperature.  On the contrary, the short time variability is characterized by redder colors when the star becomes brighter. 
This behavior can be explained in terms of increasing mass loss or/and increased contribution  from hot dust.
\begin{figure}[h]
\label{wray17}
\centering
\includegraphics[width=15cm]{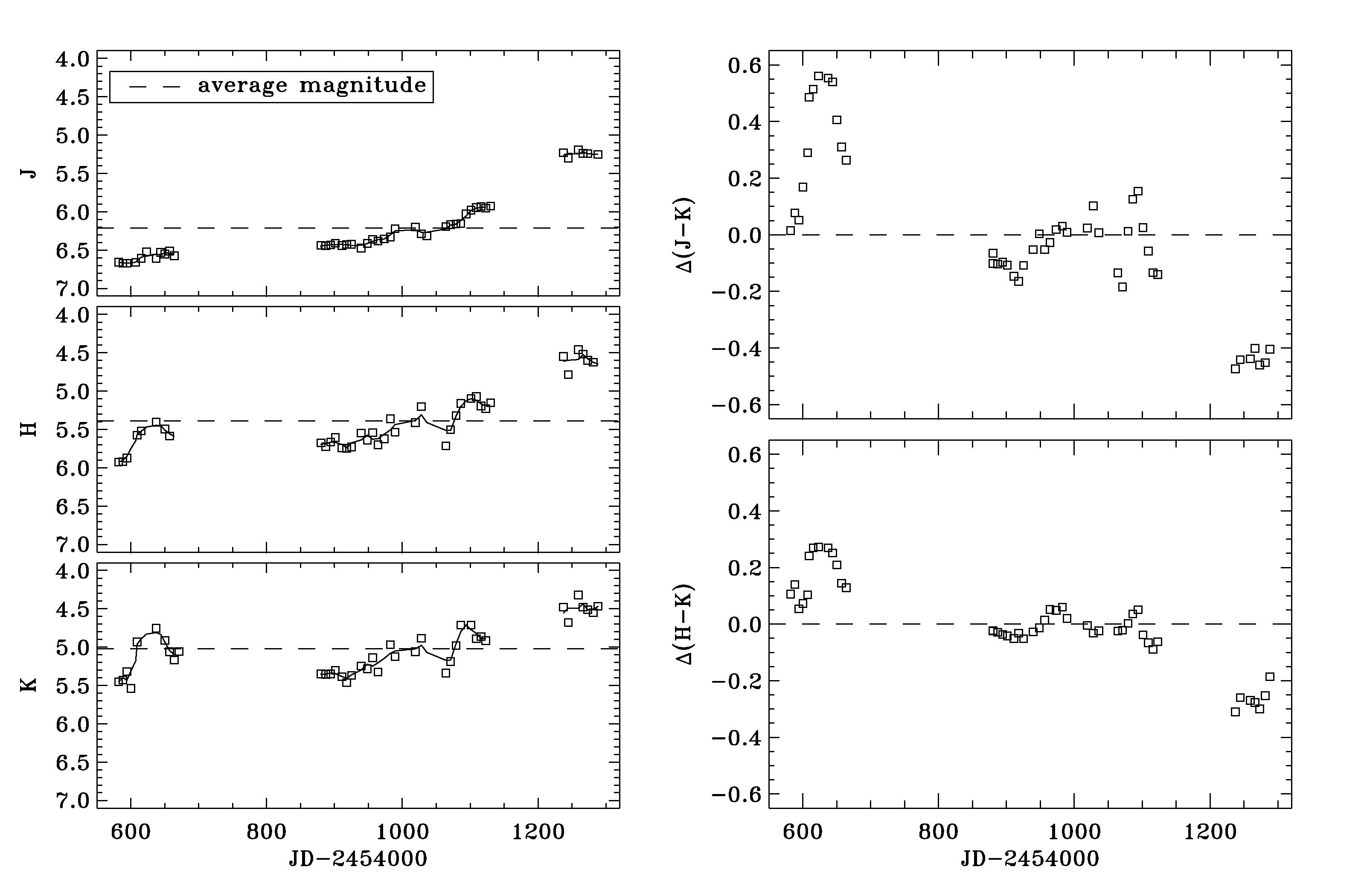}
\caption{Light and color curves for Wray~17-96. The solid lines represent the smoothed brightness.}
\end{figure}

 \subsection{LBV~1806-20}
 As shown in Fig. 3,
the observed light curves of this source show a long term rising trend, with total brightness variations having an amplitude of about 0.4~mag in all the bands from the beginning to the end of the observations.  Short term variations with amplitude no larger than 0.2-0.25~mag are observed too. 
During the long term variation, the colors remain practically constant, with a weak blueing associated to the maxima of the short term flux oscillations, suggesting that both stellar temperature and radial changing could be present.
\begin{figure}[]
\label{lbv18}
\centering
\includegraphics[width=15cm]{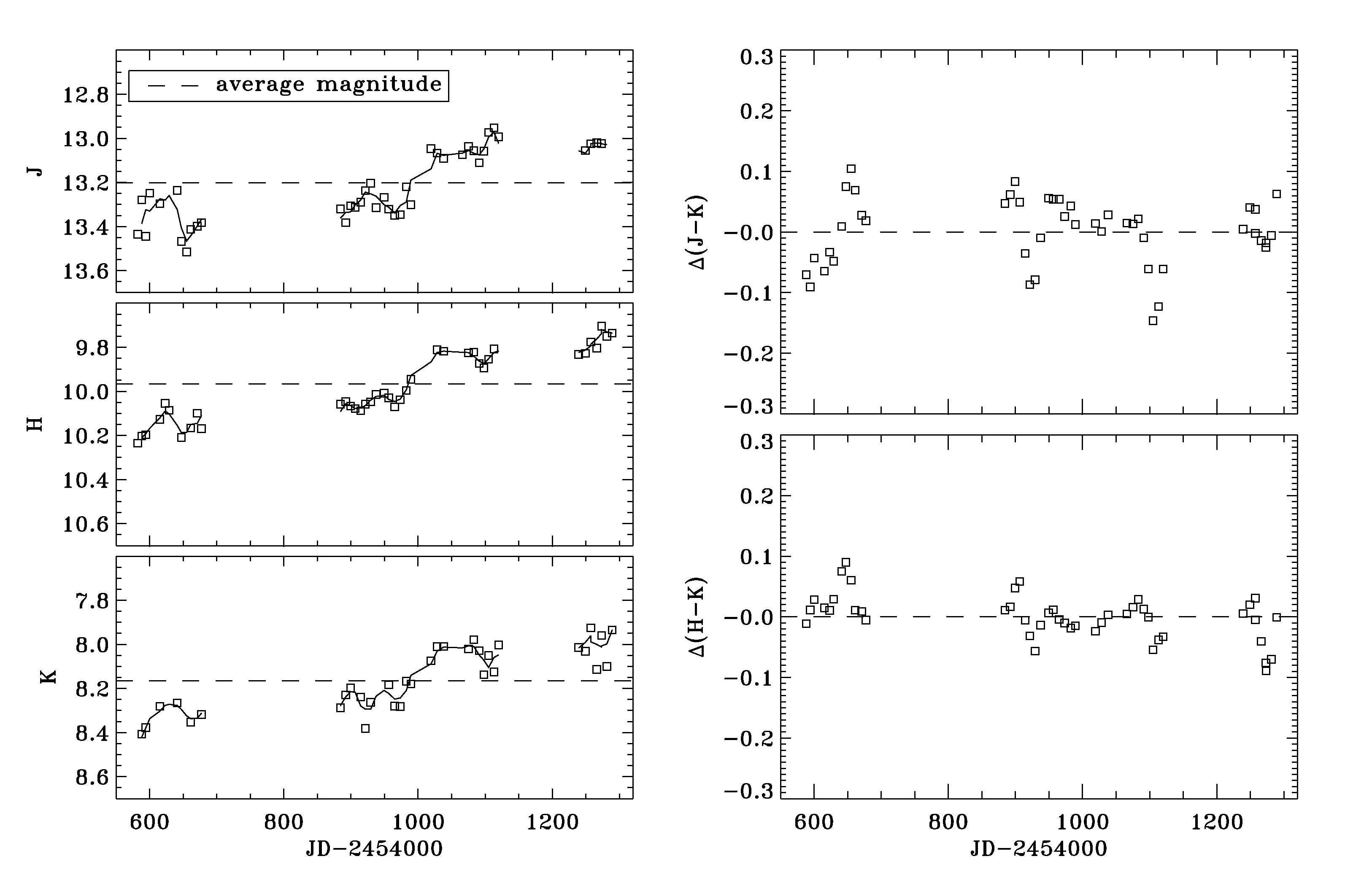}
\caption{Light and color curves for LBV~1806-20. The solid lines represent the smoothed brightness.}
\end{figure}

\section{Summary}
We are monitoring a
select group of LBV and cLBV to characterize the photometric variability changes as
a star goes through these final stages of evolution.
The aim of this monitoring program is to collect an as complete and homogenous as possible  photometric dataset which helps to 
accurately estimate the parameters of the photometric variability, as needed for a complete understanding of the properties of stars in this unstable regime.  

We observed two time scales of variation in our dataset: long term light variation covering the entire interval of our observations (about 2 years) and a small amplitude ($\Delta$mag$\approx$0.2), short time ($\approx$60 days) oscillations. We found that about 40\% of the sources in our sample are characterized by one or both such kinds of variations. Such preliminary results, support the need of a regular and homogeneous monitoring of such objects. A long term homogeneous monitoring of the objects in our sample could help to better define the distribution of the light curve parameters and to set limits on possible mechanisms generating such photometric behavior. 

The complete results of our ongoing studies will be published in Buemi et al. (2011, in preparation), where the observed light curves from the entire sample will be  discussed.\\

%
% USE A SECTION WITHOUT NUMBER FOR THE ACKNOWLEDGEMENTS
%
\section*{Acknowledgements}
Based on observations made with the REM Telescope, INAF Chile.\\
We wish to thank the REM team for technical support, and in particular Dino Fugazza, for their help in setting-up the observations.\\
We acknowledge partial financial support from PRIN-INAF 2007 and the ASI contract I/038/08/0 ``HI-GALÕ. \\

%
% BEGIN THE REFERENCE LIST WITH \beginrefer
% USE \refer BEFORE THE REFERENCES AND BEGIN A NEW PARAGRAPH AFTER THE 
% REFERENCE !
% DO NOT FORGET TO END THE LIST WITH \endrefer
%
\newpage
\footnotesize
\beginrefer
%\refer Acker A., Marcout J., Ochsenbein F., Loret M.C., 1983, A\&AS, 54, 315
%
%\refer Clark J.S., Egan M.P., Crowther P.A., Mizuno D.R., Larionov V.M., \& Arkharov A. 2003, A\&A, 412, 185
%
\refer Clark J.S., Larionov V. M., \& Arkharov A. 2005, A\&A, 435, 239.

%\refer Egan M. P., Clark J. S., Mizuno D. R., et al. 2002, ApJ, 572, 299

%\refer Eikenberry S. S., Matthews K., La Vine J. L. et al. 2004, ApJ, 616, 506
\refer Everett, M. E., Howell, S. B. 2001, PASP, 113,1428.

\refer Gilliland, R. L., Brown, T. M. 1988, PASP, 100, 754.

\refer  Humphreys R., Davidson K. 1994, PASP, 106, 1025.

\refer Kotak, R., \& Vink, J.S. 2006, A\&A, 460, L5

\refer Langer, N., Hamann, W.R., Lennon, M., Najarro, F., Pauldrach, A.W.A., Puls, J. 1994, A\&A, 290, 819.
%\refer Ratag M. A. \& Pottasch S. R., 1991 A\&AS, 91, 481
%
%\refer Umana G., Buemi C. S., Trigilio C., Hora J. L., Fazio G. G., Leto P. 2009, ApJ, 694, 697
%
%\refer Umana G., Buemi C., S., Trigilio C., \& Leto P. 2010, ApJ, in press

\refer van Genderen, A.M 2001, A\&A, 366, 508.

\endrefer         
  
\end{document}